\documentclass[12pt]{article}
\usepackage{amssymb,amsmath,latexsym, graphicx, enumerate, bigints}
\usepackage{tabularx,ragged2e,booktabs,caption}
\usepackage[font=small,labelfont=bf]{caption}
\numberwithin{figure}{subsection}

\newcommand{\di}{\frac{\delta_I}{L}}
\newcommand{\dm}{\frac{\delta_M}{L}}
\newcommand{\ds}{\frac{\delta_S}{L}}
\newcommand{\dtime}[1]{\frac{\partial \Delta#1}{\partial t}}
\newcommand{\mbf}[1]{\mathbf #1}

\newcommand{\dive}{\mathrm{div}}
\newcommand{\disp}{\displaystyle}

\begin{document}
\title{FEniCS Application to a Finite Element Quasi-Geostrophic Model.\\ \large Linear and non-linear analysis of Munk-like Solutions and Data Assimilation Implementation Through Adjoint Method.}
\author{ Maria Strazzullo\footnote {Corresponding author: maria.strazzullo92@gmail.com}, Department of Mathematics and Geoscience,\\ University of Trieste, Trieste-ITALY,\\\and Renzo Mosetti, OGS, Trieste-ITALY.}
\date{\today}
\maketitle
\abstract{\noindent The first aim of this work is to construct a Finite Elements model for quasi-geostrophic equation. This model is analyzed through FEniCS interface. As a second purpose, it shows the potential of the control theory to treat Data Assimilation large scale problems. In section one the finite element approximation of the quasi-geostrophic model is described. In section two some numerical results on different domains are shown. In section three a comparison between Navier Stokes model and quasi-geostrophic one is discussed. Last section is dedicated to a more complex data assimilation/control problem based on quasi-geostrophic equations, to test possible developments and improvements.
}

\section{Quasi-Geostrophic Model and Numerical \\Approximation}
In this section we will introduce an  unsteady quasi-geostrophic system. It describes a constant density flow moving under Earth's rotation influence. In particular we will consider the large scale wind-driven ocean circulation. For all physical theory, implications and details we refer to \cite{Crisciani, Vero1, Vero2}.
\\The equations are numerically treated by a finite element (FE) spatial discretization. Usually, the classical numerical methods used in Geophysical Fluid Dynamics are based on finite differences. This kind of approach is widespread and produces good approximated results, whereas FE methods seem to be discarded, even if they lead to very precise results, most of all where high resolution is needed: flows on coastal boundary layers and on bathymetric constraints. They fit well for post processing, data assimilation and control purposes. We refer to \cite{Alfio1, Alfio2} for FE approximation theory.

\subsection{Physical Model Strong Formulation}
Let us introduce our dynamic system on a bi-dimensional domain $\Omega$. For now $\Omega$ is not specified: in simulations we have used two different geographical domain. They will be introduced in section \ref{numerics}. Let $L = 10^6$ be the domain length scale. We can express the problem in the following way: to a given choice of a non-linear parameter $\disp \di$ and dissipative parameters $\disp \dm$ and $\disp \ds$ , find solution $\disp \psi$ that verifies, under no-slip conditions: 
$$ \dtime\psi + \Big(\di\Big)^2 \mathcal{F}(\psi, \Delta\psi)+ \frac{\partial \psi}{\partial x} = - \sin(\pi y)+ \Big(\dm\Big)^3\Delta^{2}\psi - \Big(\ds\Big) \Delta\psi,  
\footnote{It is the so called \textit{streamfunction} formulation, from which one can derive the velocity components $[u,v]$ of the current thanks to the following relations: $\disp u = -\frac{\partial\psi}{\partial y}$, $\disp v = \frac{\partial\psi}{\partial x}.$}$$ 
where $\mathcal{F}$ is the Jacobian Determinant operator\footnote{Given two differentiable functions $s(x,y)$ and $t(x,y)$, $\disp \mathcal{F}(s, t) = \frac{\partial s}{\partial x}\frac{\partial t}{\partial y} - \frac{\partial s}{\partial y}\frac{\partial s}{\partial x}. $}.
The equation is non-dimensional, so the parameters $\disp \di, \ds, \dm$ have as range $[0,1]$. A classical approach to solve this kind of problem (with a bi-Laplacian operator) is to split the equation, obtaining:

$$ \begin{cases} q = \Delta \psi &\text{ in  }\Omega, \\
 \displaystyle \frac {\partial q}{\partial t} + \Big(\di\Big)^2 \mathcal{F}(\psi, q) + \frac{\partial \psi}{\partial x} - \Big(\dm\Big)^3 \Delta q +\ds q = -\sin(\pi y)  & \text{ in  } \Omega, \\
q = 0 & \text{  on  } \partial \Omega, \\
\psi = 0 & \text{  on  } \partial \Omega.
\end{cases}$$

\subsubsection{Bathymetry}
To make the model more realistic one can add the influence of bathymetry on the equations. In the quasi-geostrophic case, it is very simple to consider this effect in the equation. Let us suppose that the sea floor can be defined by a smooth function\footnote{The value of the function must have the same order of the Rossby number.} $h: \Omega \rightarrow \mathbb{R}$. The new state equation to be examined is: 
$$ \dtime\psi + \Big(\di\Big)^2 \mathcal{F}(\psi, \Delta\psi + h(x,y))+ \frac{\partial \psi}{\partial x} = - \sin(\pi y)+ \Big(\dm\Big)^3\Delta^2\psi - \Big(\ds\Big) \Delta\psi.$$ 
To describe the Atlantic floor\footnote{Atlantic domain experiments are analyzed in subsections \ref{Atlantico} and 3.2.}, for example, one could use 
$$h(x,y) = \disp e^{-\frac{x}{\di}} + g(x,y),$$
where $g(x,y)$ is a Gaussian function representing the mid-Atlantic Ridge, whereas the exponential function represents steep continental slope.
\\ From the simulations, it turns out that the bathymetry does not influence significantly the results, so in what follows, a flat bottom is assumed.

\noindent Now we are ready to treat numerically our problem.

\subsection{Physical Model  Weak Formulation and Approximation}
This subsection describes how to find an approximated weak solution for quasi-geostrophic problem. Variational techniques are not trivial and a deep theoretical mathematical knowledge is needed to fully understand it. We leave details, information and implications to the reader: all the methods that will be mentioned are fully treated in \cite{Alfio1, Alfio2}.

\noindent Suppose $\mathbb{V}=H^1_0(\Omega)$.\footnote{$H^1_0(\Omega)$ denotes the space of square integrable functions, vanishing on $\partial \Omega$, with square integrable first derivative.} After few algebraic treatments, one gets the following formulation:  find solutions $\psi$ and $q$ in $\mathbb{V}$ such that:
$$\begin{cases}
\displaystyle \int_{\Omega}q\varphi + \int_{\Omega}\nabla  \psi\cdot \nabla  \varphi = 0 & \forall \varphi, p \in \mathbb{V},\\
\displaystyle
\int_{\Omega}\frac{\partial q}{\partial t}p +
 \Big(\di\Big)^2  \int_{\Omega}(\psi \nabla q \times \mathbf{\hat k })\cdot \nabla p +
 \int_{\Omega}\frac{\partial\psi}{\partial x}p + \Big(\dm\Big)^3 \int_{\Omega}\nabla q \cdot \nabla p + \ds \int_{\Omega}q p = f&  \forall \varphi, p \in \mathbb{V},\\
\end{cases}$$
where $f = -\sin(\pi y)$. 
\\
We endow $\mathbb{V}$ with the inner product: $$(v,w)_\mathbb{V} = \displaystyle \int_{\Omega}\nabla v \cdot \nabla w \hspace{1.0cm} \forall v,w \in \mathbb{V}. $$
Now a discrete approximation $\mathbb{V}_\delta \subset  \mathbb{V}$ is considered. The discretization $\mathbb{V}_\delta$ has been built as a standard FE approximation. The discretized system version is:

$$\begin{cases}
\displaystyle \int_{\Omega}q_\delta\varphi_\delta + \int_{\Omega}\nabla  \psi_\delta\cdot \nabla  \varphi_\delta = 0 & \forall \varphi_\delta, p_\delta \in \mathbb{V_\delta},\\
\displaystyle
\int_{\Omega}\frac{\partial q_\delta}{\partial t}p_\delta +
 \disp \Big(\di\Big)^2  \int_{\Omega}(\psi_\delta \nabla q_\delta \times \mathbf{\hat k })\cdot \nabla p_\delta
 \disp + \int_{\Omega}\frac{\partial\psi_\delta}{\partial x}p_\delta + \\ \qquad \qquad \qquad \qquad
 \disp +  \Big(\dm\Big)^3 \int_{\Omega}\nabla  q_\delta \cdot \nabla p_\delta + \ds \int_{\Omega} q_\delta p_\delta = f_\delta &  \forall \varphi_\delta, p_\delta \in \mathbb{V}_\delta.

\end{cases}$$

\noindent The solutions $q_\delta$ and $\psi_\delta$ will be good approximations of $q$ and $\psi$ in the inner product norm.
\\ For the computational implementation we used \textit{FEniCS} interface (\textbf{https://fenicsproject.org/}), exploiting  
\textit{DOLFIN} library (\textbf{https://fenicsproject.org/about/components.html}).

\section{Numerical Results}
\label{numerics}
This section aims at showing some numerical results in two different domains: the unit square and a North Atlantic Ocean grid\footnote{In order to build the North Atlantic mesh, we exploited \textit{Freefem++} (\textbf{http://www.freefem.org/}) and \textit{gmsh} (\textbf{http://gmsh.info/}).}. The square is used to demonstrate equation stability. From the results we can also deduce that FE model is comparable to finite differences one. After this analysis, we implemented simulations on a more realistic coastal grid, representing the Northern part of Atlantic Ocean.
We underline that time scale of the model is of the order of $O(10^8)$ seconds.

\subsection{Simulation on a Square Domain}
As previously introduced, the square $\Omega = [0,1] \times [0,1]$ is considered to be the fluid domain. It represents a wide Ocean surface portion. We ran our model with a time increment $dt = 10$ (almost 4 months). Simulation plots at different time steps are shown in the following lines. One can observe that at final time the system reach the steady solution profile. Results have been compared to streamfunction profile presented in \cite{Crisciani}. In all the experiments $\disp \ds = 0$ is considered.  We analysed three cases representing three typical geophysical Ocean behaviours: 
\begin{itemize}
\item linear case, with $\disp \di = 0$ and $\disp \dm = 7\cdot 10^{-2}$,
\item low non-linear case, with $\disp \di = \dm = 7\cdot 10^{-2}$,
\item highly non-linear case, with $\disp \di = 7\cdot 10^{-2}$ and $\disp \dm = 7\cdot 10^{-3}$.
\end{itemize}
The number of iterations needed to the convergence of the solver is reported. It will be indicated by IN. Time evolution is given by implicit Euler method with null initial condition. \\
In all the plots cold colors are used to represent the low values of the solution, while hot colors represent high values.
\noindent Let us focus on the highly non-linear model:  final stationary convergence has few variations from \cite{Crisciani}. However, our FE solution is physically meaningful and coherent.  The results are normalize with respect to maximum values. The simulations can be compared with the Munk-like model presented in \cite{Crisciani}.


\begin{figure}[h]
\centering
\includegraphics[scale = 0.6]{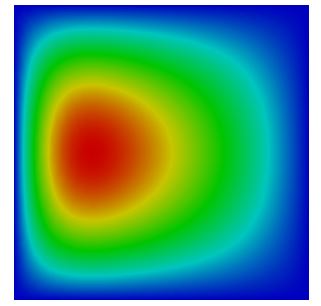} \includegraphics[scale = 0.6005]{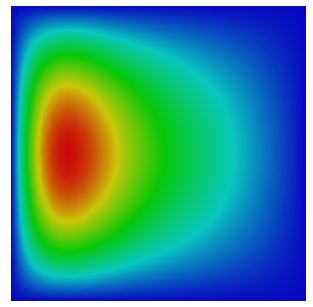} \includegraphics[scale = 0.6007]{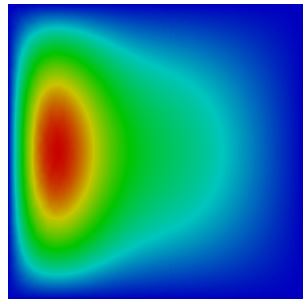}
\caption{Linear model: left t = 10 and IN = 1,  center t = 30 and IN = 2, right t = 60 and IN = 2 (stationary convergence).}
\end{figure}

\begin{figure}[h!]
\centering
\includegraphics[scale = 0.6]{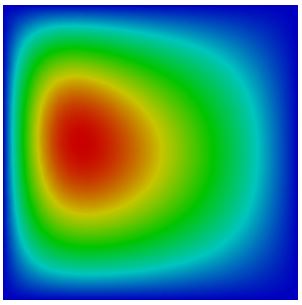} \includegraphics[scale = 0.6005]{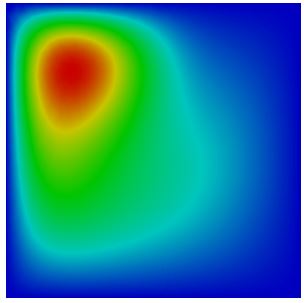} \includegraphics[scale = 0.6007]{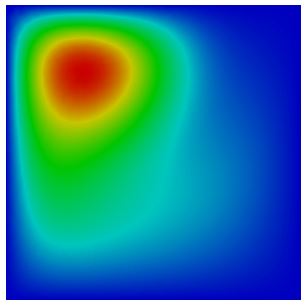}
\caption{Low non-linear model: left t = 10 and IN = 5, center t = 60 and IN = 4, right t = 130 and IN = 4 (stationary convergence).}
\end{figure}

\begin{figure}[h!]
\centering
\includegraphics[scale = 0.6]{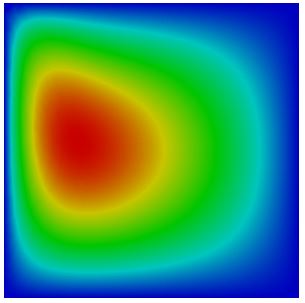} \includegraphics[scale = 0.6006]{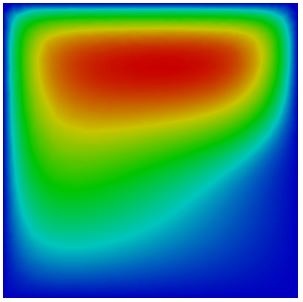} \includegraphics[scale = 0.6007]{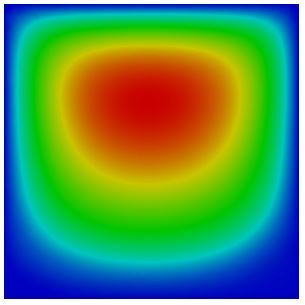}
\caption{Highly non-linear model: left t = 10 and IN = 6,  center t = 60 and IN = 7, right t = 130 and IN = 5 (stationary convergence).}
\end{figure}

\newpage 
\noindent As one can observe, the  main difference between linear and non-linear solutions is that the linear system provides a thickening of the current on Western boundary, whereas non-linearity leads the solution to a deviation to North-East.
\subsection{Simulation on North Atlantic Ocean}
\label{Atlantico}
In this subsection simulations are on a more complex and realistic domain. We choose to analyse the streamfunction dynamics on the North Atlantic Ocean. We assume no-slip conditions all around the boundaries and $\disp \ds = 0$ is considered. The time step chosen is $dt = 10$. We simulated until stationary convergence. Three cases are studied:
 \begin{itemize}
\item linear case, with $\disp \di = 0$ and $\disp \dm = 7\cdot 10^{-2}$,
\item low non-linear case, with $\disp \di = \dm = 7\cdot 10^{-2}$,
\item highly non-linear case, with $\disp \di = 7\cdot 10^{-2}$ and $\disp \dm = 7\cdot 10^{-3}$.
\end{itemize}

\noindent The time evolution and IN index are reported in the following plots. The initial condition for implicit Euler method solver is the null fields. The color map for the streamfunction is as before. As in the previous example, a more pronounced North-East flow from North America to Europe is observed when a higher value of $\disp \di$ is assumed. The results are normalized with respect to maximum of the solution.
 
\vspace{1cm}

\begin{figure}[h!]
\centering
\includegraphics[scale = 0.6]{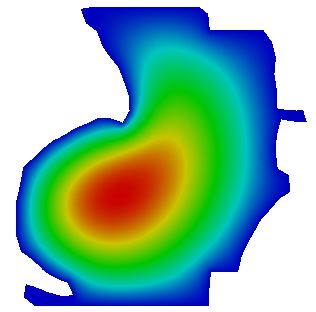} \includegraphics[scale = 0.6]{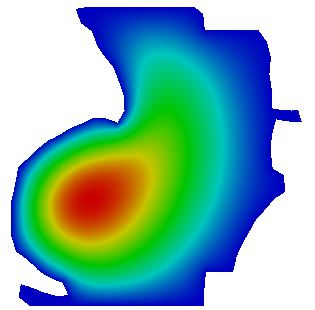}
\includegraphics[scale = 0.6]{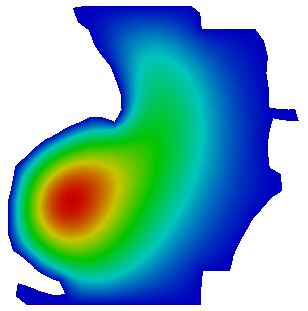}
\caption{Linear model: left t = 10 and IN = 1,  center t = 30 and IN = 1, right t = 100 and IN = 1 (stationary convergence).}
\end{figure}

\begin{figure}[h!]
\centering
\includegraphics[scale = 0.6]{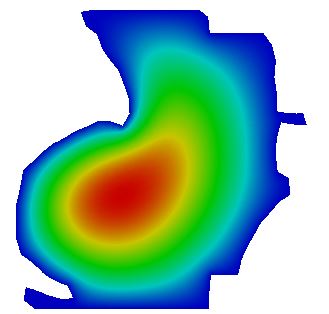} \includegraphics[scale = 0.6]{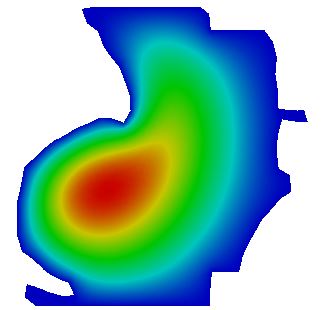} \includegraphics[scale = 0.6]{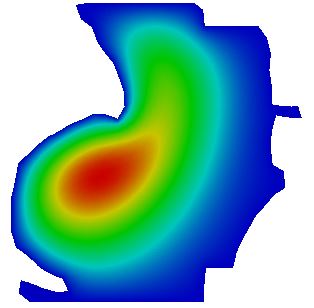}
\caption{Low non-linear model: left t = 10 and IN = 1, center t = 30 and IN = 4, right t = 100 and IN = 4 (stationary convergence).}
\end{figure}

\begin{figure}[h!]
\centering
\includegraphics[scale = 0.6]{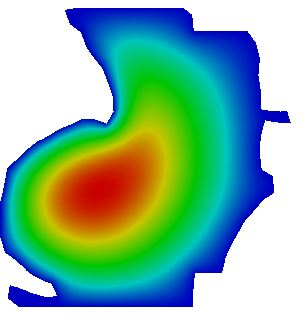} \includegraphics[scale = 0.6]{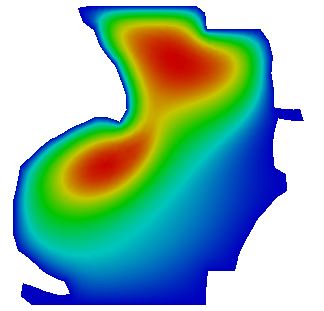} \includegraphics[scale = 0.6]{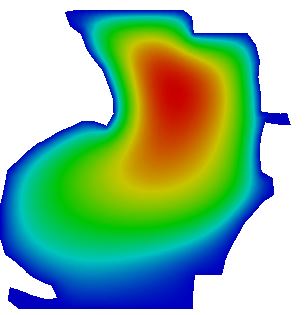}
\caption{Highly non-linear model: left t = 10 and IN = 1, center t = 60 and IN = 7, right t = 100 and IN = 5 (stationary convergence).}
\end{figure}

\newpage
\section{Why not Using Navier Stokes Equations?}
After this analysis one can move to a more general model for flow dynamics: Navier Stokes equations. Quasi-Geostrophic equations derive directly from them, after some scaling procedures. Quasi-geostrophic system could be seen as an approximation, a simplified model that explain large scale circulation, reliably. By adopting the primitive equations, a more precise description of the process could be, in principle, obtained. But is it convenient? This section describes Navier Stokes general model, its numerical results and how it is instable under high non-linearity hypothesis.

\subsection{Strong and Weak Formulation}
Let us introduce incompressible Navier Stokes model in a bi-dimensional large scale domain $\Omega$, governed by the following system in strong form:
$$\begin{cases}
\displaystyle \frac{\partial \mbf{u}}{\partial t} -  \Big(\dm\Big)^3 \Delta \mbf u + \Big(\di\Big)^2 (\mbf u \cdot \nabla)\mbf u 
+   \Big(\ds\Big)\mbf u
+ \nabla p = \mbf f, \\
\mathrm{div}(\mbf u) = 0, 
\end{cases}$$
where $\mbf u = [u,v]$ and $p$ are flow velocity and pressure, respectively, and $\disp \mbf f = [- \frac 1 \pi \cos(\pi y)]\footnote{This value is linked to $-\sin(\pi y)$: for details see \cite{Crisciani}.}, 0 ]$ is the wind stress. First of all, we are not considering the streamfunction anymore. The current itself is our output. After a little algebra, exploiting no-slip conditions for velocity, the following weak formulation for $\mbf u$ and p is reached: find the two components $[u,v]$ in $\mathbb {V} = H^1_0(\Omega)\times H^1_0(\Omega)$ and $p \in \mathbb {P} = L_0^2(\Omega)$\footnote{$ L_0^2(\Omega) = \left \{ p \in L^2 (\Omega) \; : \; \displaystyle \int_{\Omega} p = 0\right \}$, where $L^2(\Omega)$ is the space of square integrable functions.} such that:

$$\begin{cases}
\displaystyle
\int_{\Omega} \frac {\partial \mbf u}{\partial t}\cdot \mbf w + 
\Big(\dm\Big)^3 \int_{\Omega}\nabla \mbf u \cdot \nabla \mbf w +
 \Big(\di\Big)^2 \int_{\Omega}(\mbf u \cdot \nabla)\mbf u \cdot \mbf w +
\displaystyle
 \Big(\ds\Big) \int_{\Omega }\mbf u \cdot \mbf w -  \int_{\Omega }\mbf f \cdot \mbf w 
- \int_{\Omega} \dive(\mbf w)p= 0 \\
\displaystyle \int_{\Omega}\dive(\mbf u)s = 0, 

\end{cases}$$

\noindent for all $\mbf w \in \mathbb{V}$ and $s \in \mathbb {P}$.
In this new framework, we are not considering Coriolis' effect. In the strong formulation, it can be recovered, in the $\beta$ plane, adding 
$$
\begin{aligned}
-(1+y)v & \hspace {0.5cm}\text{for the first equation},\\
(1+y)u & \hspace {0.5cm}\text{for the second equation},
\end{aligned}
$$
that is, in weak formulation, for all $\mbf w = [w_1, w_2] \in\mathbb{V}$:
$$
\begin{aligned}
\displaystyle - \int_{\Omega} (1+y)vw_1 & \hspace {0.5cm}\text{for the first equation},\\
\displaystyle \int_{\Omega} (1+y)uw_2 & \hspace {0.5cm}\text{for the second equation}.
\end{aligned}
$$

\noindent Now, after a finite element approximation, the model is ready to be numerically treated. 

\subsection{Simulation on a square domain}
We considered $\Omega = [0,1]\times [0,1]$, to compare results with quasi-geostrophic model. For the same reason, $\disp \ds = 0$ is chosen. A time step $dt = 10$ is taken and we ran our simulation exploiting implicit Euler until stationary convergence, from a null initial condition.  What we expect is a gyre thickening to the Western boundary for the linear case, whereas a gyre moving to North-East for the non-linear case. We assume no-slip conditions for the velocity field.

 \begin{itemize}
\item linear case, with $\disp \di = 0$ and $\disp \dm = 7\cdot 10^{-2}$,
\item low non-linear case, with $\disp \di = \dm = 7\cdot 10^{-2}$,
\end{itemize}

\noindent The color map is the usual cold-hot low-high. Time steps and IN index are reported. Numerical results are normalized with respect to the maximum of the solution.

\begin{figure}[h!]
\centering
\includegraphics[scale = 0.6]{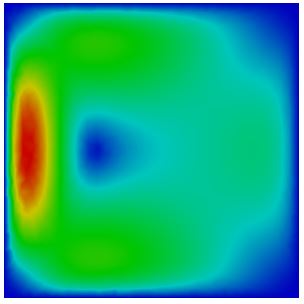} \includegraphics[scale = 0.6005]{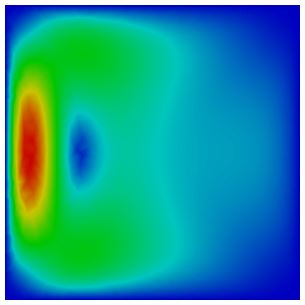} \includegraphics[scale = 0.6007]{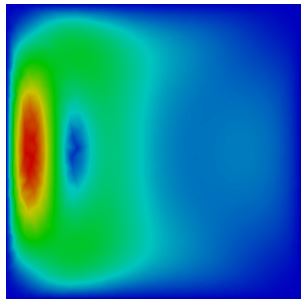}
\caption{Linear Navier Stokes model: left t = 10 and IN = 2,  center t = 30 and IN = 2, right t = 60 and IN = 2 (stationary convergence).}
\end{figure}

\begin{figure}[h!]
\centering
\includegraphics[scale = 0.6]{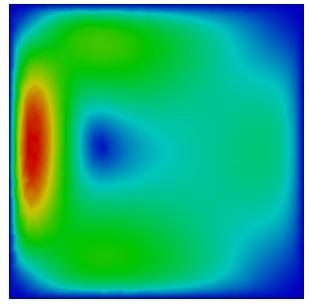} \includegraphics[scale = 0.6005]{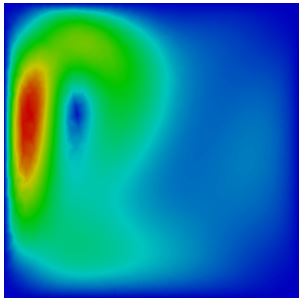} \includegraphics[scale = 0.6007]{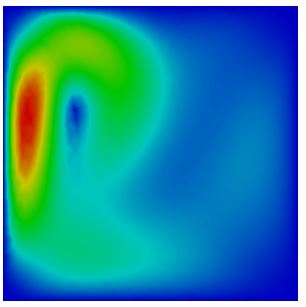}
\caption{Low non-linear Navier Stokes model: left t = 10 and IN = 4,  center  t = 60 and IN = 8, right  t = 100 and IN = 3 (stationary convergence).}
\end{figure}
\newpage
\noindent Now, let us analyse the highly non-linear configuration with  with $ \displaystyle\di = 7\cdot 10^{-2}$ and $\displaystyle \dm = 7\cdot 10^{-3}$. Navier Stokes equations are very instable. Considering $dt = 5$ we have the following result for t = 10.

\begin{figure}[h!]
\centering
\includegraphics[scale = 0.6]{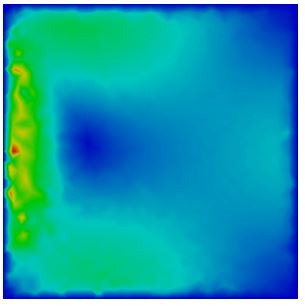} 
\caption{Highly non-linear Navier Stokes model: instability.}
\end{figure}

\noindent This plot is not surprising if seen in Reynold number terms. Indeed $$\disp \frac {1}{\text{Re}}=\Big ( \di \Big)  ^{-2}\Big (\dm \Big) ^3= O(10^{-5}).$$  So, for high values of non-linearity, Navier Stokes is an instable model and could  not represent large scale wind-ocean circulation in an efficient and reliable way. 

\subsection{Simulation of the North Atlantic Ocean}
One can consider Navier Stokes model to describe circulation in a realistic domain, like the North Atlantic Ocean. The general equations, in principle, are able to better capture flow dynamics, but we have to face a very huge stability issue in the non-linear case, as in the square domain experiment. Using the North Atlantic domain, the algorithm does not converge for $
\displaystyle \di \gg\dm$. This is the reason why highly non-linear case is not reported.  We can only show results of: 
 \begin{itemize}
\item the linear case, with $\disp \di = 0$ and $\disp \dm = 7\cdot 10^{-2}$,
\item the low non-linear case, with $\disp \di = \dm = 7\cdot 10^{-2}$.
\end{itemize}
\noindent  As usual: $\disp\ds = 0, dt = 10$. Time simulation is based on implicit Euler method, starting from a null fields. No-slip boundary conditions are assumed.
\begin{figure}[h!]
\centering
\includegraphics[scale = 0.6]{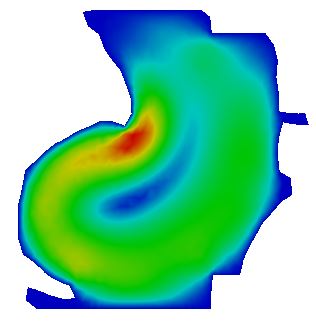} \includegraphics[scale = 0.6]{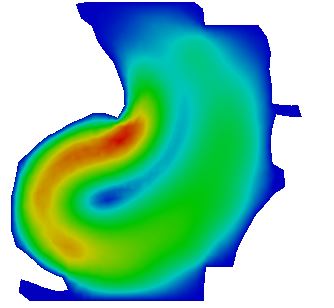} \includegraphics[scale = 0.6]{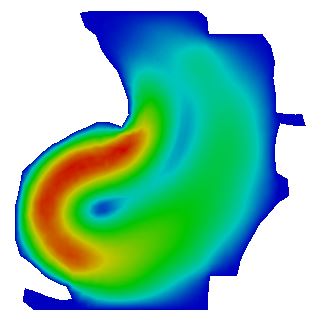}
\caption{Linear Navier Stokes model: left t = 10 and IN = 2,  center t = 30 and \\IN = 2, right t = 60 and IN = 2 (stationary convergence).}
\end{figure}

\begin{figure}[h!]
\centering
\includegraphics[scale = 0.6]{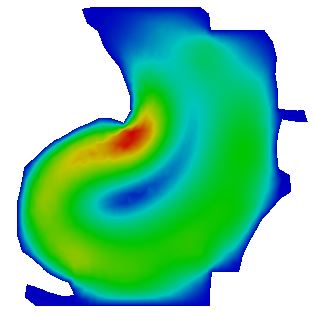} \includegraphics[scale = 0.6]{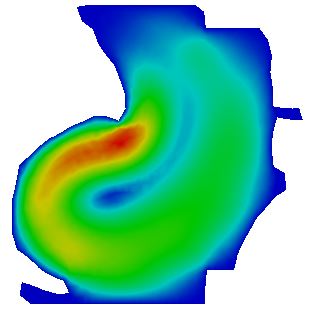} \includegraphics[scale = 0.6]{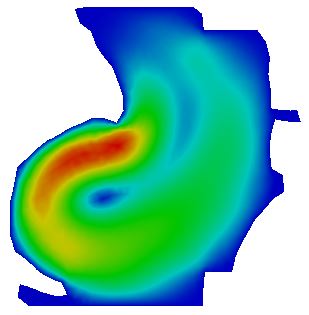}
\caption{Low non-linear Navier Stokes model: left t = 10 and IN = 4,  center \\ t = 60 and IN = 4, right  t = 100 and IN = 4 (stationary convergence).}
\end{figure}

\noindent As we expected, in linear case we have a thickening phenomenon on American East coast. In low non-linear case one can see how the current moves towards North-East.

\subsection{Some Comments}
Let us draw some conclusions on the advantages and disadventages of using the Navier Stokes equations. Indisputably, they are a complete and meaningful tool to describe Ocean dynamics. Although, computationally speaking, they are very difficult to treat. Quasi-geostrophic model can afford better the non-linear issues.
\\ Note that Navier Stokes model and quasi-geostrophic model are comparable in linear and low non-linear cases: the IN index are similar. This underlines how Navier Stokes equations can be an important instrument to study the examples presented. In the case of high non-linearity, stabilized FE methods are needed \cite{Mitri, Dani}.

\section{Data Assimilation}
Data Assimilation is a technique that allows to be more reliable in forecasting models. How can we chance model features in order to reach a desired output? What is the forcing term needed in order to have a specific solution? In our case we have built an inverse problem on wind stress\footnote{The idea is based on the scenarios presented in \cite{Vento}.}. What is the wind intensity that allows our output to be similar to a predicted scenario? This fits perfectly in Data Assimilation concept: try to incorporate data information in our basic model, changing it, in order to be more reliable. 
\\Data Assimilation is based on optimal control problems, faced through adjoint method. The control theory fundamentals are not taken into account in this work, but they are fully treated in \cite{Alfio2, Lions, Ronald}. In \cite{Eli} adjoint method is specifically applied to ocean circulation models. 
\subsection{A Specific Example}
Let us consider the square domain $\Omega = [0,1]\times [0,1]$ and a quasi-geostrophic description of our dynamics. We are interested to assimilate and control terms on the West boundary layer. 
\\Suppose that we have some dataset $D_1, \dots, D_n$ (representing more powerful simulation data, experimental data, different scenarios...) at some time values $t_1, \dots, t_n$. The interpolation of $D_1, \dots, D_n$ on the considered domain, gives $\bar \psi_1, \dots, \bar \psi_n$ functions. Let $\psi_i$ be the quasi-geostrophic solution at time $t_i$ for $i = 1, \dots, n$. Our goal is to modify $\psi_i$ in order to fit  $\bar \psi_i$. The only way to obtain this result is to change parameters or forcing term. In our case we decided to analyse the wind forcing action needed to reach the desired state solution. For this purpose, we exploited Lagrangian adjoint method.
Let us define the following objective functional for all $i$: 
$$J(\psi_i, f) = \frac 1 2 \displaystyle \int_{\Omega}(\psi_i - \bar \psi_i)^2 + \frac \alpha 2 \int_{\Omega}f^2,$$
where $\alpha$ is a penalization term. The aim of the problem is to minimize the functional, constrained to the state equations: 
$$\underset{(\psi_i,f)}{\text{min}}J(\psi_i, f)
\hspace{1cm}
\text{subjected to } 
\hspace{1cm}
\mathcal {E}(\psi_i, f)=0,$$
where $\mathcal {E}(\psi_i, f)=0$ represents our state dynamic.
\\Supposing that, at least locally, for each $f$ a unique $\psi_i$ exists, so we can write our solution as $\psi_i(f)$. Under some mathematical hypotheses, the minimization problem is equivalent to find:
$$\underset{f}{\text{min }}J(\psi_i(f), f)
\hspace{1cm}
\text{subjected to } 
\hspace{1cm}
\mathcal {E}(\psi_i(f),f)=0.$$
Let $\hat J(f) := J(\psi_i(f), f)$ be the so called reduced functional. Now we have all the ingredients to solve our issue, thanks to Lagrangian formalism and adjoint method.

\subsection{Numerical results}
In the following subsection we are going to show some basic results, by applying an adjoint approach to geophysical flow modeling.\\ First of all, let us clarify the specifics of the experiment. Simulation ran until final time $T=90$. We wanted to enrich our model thanks to given data\footnote{The data have obtained from a finite differences MATLAB simulation.} at time $t_1 = 30, \; t_2 = 60, \; t_3 = 90.$ 
These data were imported in \textit{FEniCS} interface in order to use \textit{dolfin-adjoint} (\textbf{www.dolfin-adjoint.org}) library: it is able to derive the discrete adjoint model from a state constrained minimization problem. For the use of this optimization library we refer to \cite{Dolfin}. \\ At each assimilation time a $f_i^{opt}$ is found and promptly substituted to the old $f$ value. The parametric setting is the following: $\disp \ds = 0$ and $\disp \di = \dm = 2 \cdot 10^{-2}.$
\\As usual we normalized the solutions with respect to the maximum value of the solution. As initial wind stress $f = -\sin(\pi y)$ is considered. From the results one can notice that the original solution attempts to reach data information as we expected.

\begin{figure}[h!]
\centering
\includegraphics{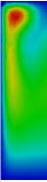}\hspace{3cm} \includegraphics[scale = 1]{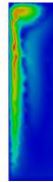} \hspace{3cm} \includegraphics[scale = 1]{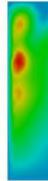} \\
\caption{Streamfunction final time assimilation: left original simulation,  center assimilated simulation, right interpolated data.}
\end{figure}

\newpage
\section*{Conclusions and Perspectives}
This work underlines that Finite Element approach is comparable to classical finite differences methods. Results and simulations allow to conclude that FE methods could be a valid option to analyse quasi-geostrophic problems.
This report has a strong educational purpose and could be an inspiration for other studies.
\\ The first improvement could involve the parametric study of the problem: one can simulate for different values of the parameters and compare the solutions. From the mathematical point of view, parametric problems can be faced trough particular numerical methods, like reduced order methods \cite{Rozza}. These are faster than standard Finite Element approaches. They are very useful when one has to manage a huge amount of parameters. Parametric optimal control problems could be a very powerful instruments to study either geophysical either environmental phenomena and they totally fit in Atmospheric and Oceanographic science.

\section*{Acknowledgements} 
First of all, I want to thank the \textit{Instituto Nazionale di Oceanografia e Geofisica Sperimentale} (OGS), Trieste, Italy, that allow me to write this report about the results obtained during my period of internship. \\ In particular, in the development of the ideas presented in this work, I would like to acknowledge the support of Professor Renzo Mosetti, OGS. \\ For the technical and computational advices, I thank Dr. Francesco Ballarin, SISSA. \\ I would also show my gratitude to Professor Gianluigi Rozza, SISSA, who introduced me to the control theory applied in marine sciences and incited me in this internship experience.

\end{document}